\documentstyle[twocolumn,prb,aps]{revtex}
\begin{document}

\draft
\title{Vortex fluctuations in superconducting $La_{2-x}Sr_xCuO_{4+\delta }$}

\author{Yung M. Huh and D. K. Finnemore}
\address{Ames Laboratory, USDOE and Department of Physics and Astronomy,\\
Iowa State University, Ames, IA 50011}

\date{\today}
\maketitle
\begin{abstract}
Vortex fluctuations in the $La_{2-x}Sr_{x}CuO_{4+\delta }$ system have been studied 
as a function of magnetic field, temperature and carrier concentration in order to determine the 
dimensionality of the fluctuations.  For a $x=0.10$ sample, there is a unique crossing-temperature on the 
magnetization vs. temperature plots for all magnetic fields up to $7~T$, and the data scale very 
well with 2D fluctuation theory.  At lower $x$-values where $H_{c2}$ is much smaller, 
there are two well defined crossing points, 
one at low fields (typically less than $1~T$) and another at high fields (typically $3-7~T$).  A fit of the data to fluctuation theory shows that the low field crossing data scale as 2D fluctuations and the 
high field crossing data scale as 3D fluctuations.  It would appear that as the magnetic field 
approaches $H_{c2}$, there is a $2D$ to $3D$ cross-over where  
the low field 2D pancake vortex structure transforms into a 3D vortex structure. 

\end{abstract}

\pacs{74.30.Ci., 74.30.Ek, 74.40.+k, 74.60.-w, 74.70.Vy}

\section{Introduction}

The existence of vortex fluctuations close to the superconducting transition temperature, $T_c$, 
in the high temperate superconductors was reported by Kes 
and coworkers,\cite {1} and theoretical work by Bulaevskii and coworkers\cite{2}
indicated that entropy terms associated with moving vortices could provide a way 
to understand the unique crossing-point in the magnetization vs. 
temperature plots that are observed.  Early work\cite {2} focussed on the flucutation of $2D$ pancake 
vortices in materials with a very high anisotropy ratio of the effective mass of the electrons, 
$\gamma = \sqrt {m_c/m_{ab}}$.    Subsequent experimental and theoretical work by Tesanovic and coworkers\cite{3}and Welp and coworkers\cite {4} indicated that the similar crossing-point 
effects also occur in materials with smaller $\gamma $-values, but the fluctuations may have  
$3D$ character.  In the case of $2D$ fluctuations, the magnetization scales to a universal curve 
when the data are plotted as $4\pi M/(TH)^{1/2}$ vs. $[T-T_c(H)]/(TH)^{1/2}$.  In the case of 
$3D$ fluctuations, the magnetizations scales to a universal curve when the data are plotted as 
 $4\pi M/(TH)^{2/3}$ vs. $[T-T_c(H)]/(TH)^{2/3}$.  These scaling laws provide an easy test 
to determine whether the fluctuations have a $2D$ or $3D$ character.
 
A recent study of vortex fluctuations in an underdoped $YBa_2Cu_3O_{6.5 }$ sample  
by Poddar and coworkers\cite {5}has  
indicated that there were two distinct crossing points for different magnetic field 
ranges.  Data from $0.2~T$ to $0.75~T$ show  a 
crossing point at $45.2~K$ and $3D$ scaling.  Data from $1.5~T$ to $3.5~T$  show a crossing point at $42.8~K$ and $2D$ scaling.   Theoretical analysis to 
explain these data\cite {6} uses a Josephson interlayer coupling  Hamiltonian that gives $2D$ 
behavior if the ratio of the c-axis coherence distance to copper 
oxide plane spacing, $\xi _c/ d$, is much less than 1 and $3D$ behavior if $\xi _c/d$ is substantially 
more than one.  For the Poddar and coworkers sample,\cite {5} the ratio of the crossing temperature to transition temperature is 
$T_{3D}^*/T_{co}=[43.4~K/45.15~K]=0.961
$ and $T_{2D}^*/T_{co}=[42.8~K/45.15~K]=0.948$.

Several types of behavior seem to occur.  For a highly anisotropic superconductor like $Bi_2Sr_2Ca_1Cu_2O_{8+\delta }$\cite {1} 
where the 
critical field at zero temperature, $H_{c2}(0)$, is much larger than the measuring field, the 
magnetization vs. temperature curves, $4\pi M~vs.~T$, have a single well defined crossing-point, $T^*$ and 
the fluctuations have $2D$ character.  For a more isotropic superconductor like optimally 
doped $YBa_2Cu_3O_{6.95}$,\cite {4} there also is a single well defined $T^*$ and 
the fluctuations have $3D$ character.  For underdoped $YBa_2Cu_3O_{6.65}$,\cite {5} 
there can be more than one crossing-point and a cross-over from $2D$ to $3D$ behavior.
The purpose to the work reported here is to systematically study the change in fluctuations 
that occur as the doping level is decreased in  $La_{2-x}Sr_xCuO_{4+\delta }$ to look 
for a systematic transformtion from $3D$ to $2D$ character and to look for the values of 
reduced magnetic field, $H/H_{c2}(T=0)$ where this cross-over occurs.

\section{Experiment}

The samples used in this work are the same grain-aligned powders and single crystals that were used for measurments of the thermodynamic critical field.\cite {7}  Powders were ground to a particle size 
less than $20~\mu m$, placed in a low viscosity epoxy and aligned in a field of $8~T$.  After the 
epoxy hardened,  X-ray rocking curves showed a full width at half maximum for the (008) peak 
of $5^\circ $.  Magnetization studies were made with a Quantum Designs magnetometer.\cite {7}
In all of the scaling analyses presented here, the magnetic field dependent transition temperature, 
$T_c(H)$ is taken from the Hao-Clem analyses of  thermodynamic data presented 
previously.\cite {7,8}  A full discussion comparing different assumptions for $T_c(H)$ is 
given elsewhere.\cite {9}  

\section{Results and discussion}

As reported earlier,\cite {8} the magnetization vs temperature curves for the $x=0.10$ 
single crystal with $T_{co}=26.8~K$ showed a 
single crossing point at $T^*=22.0~K$ and $4\pi M^*=-1.13~G$ for all data from $1.0-7.0~T$. 
No extensive study was made at lower fields.  Fits of these data on a 
 $4\pi M/(TH)^{1/2}$ vs. $[T-T_c(H)]/(TH)^{1/2}$ plot show $2D$ scaling over this 
entire field range.  To discuss the case where the crossings occur at different temperatures,  we define a new variable
$T_{cr}$ as the temperature where two successive $4\pi M~vs.~T$
curves cross and plot
$T_{cr}$ as a function of magnetic field to show how
the crossing point changes with field.

Data for a sample close to optimum doping at $x=0.143$ are presented in Fig. 1.  The data from 
$0.50~T$ to $0.95~T$ cross at about $36.8~K$ and the data from $2~T$ to $7~T$ cross at 
about $35.4~K$, in a manner similar to the $YBa_2Cu_3O_{6.65}$ data reported by 
Poddar.\cite {5} These temperatures are fairly close to one another, but they are 
easily resolved.   If data for these two samples are fit to both $2D$ and $3D$ scaling, the $3D$ scaling 
gives a significantly better fit than the $2D$ scaling for both plateaus.  The data on Fig. 2 
for the two samples close to optimum doping, $x=0.156$ (open stars) and $x=0.143$, 
(solid pentagons) are very similar.  Within the theory of Rosenstein et al.\cite {6} the relevant 
quantity to determine the scaling dimensionality is $b=H/H_{c2}(0)$.  For these samples close to optimum 
doping, the upper critical field, $H_{c2}(0)$, is approximately $33~T$, so the measurments up 
to $7~T$ never probe the region close to $H_{c2}(0)$ so the $b=H/H_{c2}(0)$ is small.  
.
Data for the $x=0.10$ single crystal shown by the solid triangles in Fig. 2, show that $T_{cr}$ 
drops smoothly with field and forms a plateau at $T^*=22.0~K$ as reported earlier.\cite {8} 
For this sample, fits to both $2D$ and $3D$ scaling give a rather good fit to $2D$ 
scaling\cite {8}, but a poor fit to $3D$ scaling.  For this $x=0.10$ sample,  $H_{c2}(0)=34~T$ is much higher 
than the measuring field of $7~T$, so the data never approach the upper critical field.  
  The difference between the $x=0.143$ sample and the $x=0.10$ sample is that the optimum doped 
sample obeys $3D$ scaling and the underdoped sample obeys $2D$ scaling.  This would imply 
that the c-axis coherence distance to copper oxide plane spacing, $\xi _c/d$, 
is always less than one for the magnetic fields measured.

Data for an $x=0.117$ sample are shown by the solid diamonds on Fig.2.  The $T_{cr}$ 
data are similar to the $x=0.10$ sample and show $2D$ behavior over the range measured.  
For this sample, $H_{c2}(0)=32~T$, so the measurements up to $7~T$ 
do not probe anywhere close to $H_{c2}(0)$. 

In the far underdoped region, data for the $x=0.081$ and $x=0.070$ samples 
differ from the above cases in that the 
respective upper critical fields of $H_{c2}(0)=11~T$ and $H_{c2}(0)=6~T$ are much closer to 
the top measuring field of $7~T$.  The $4\pi M~vs.~T$ data for the $x=0.081$ sample 
presented  in Fig. 3 show a low field crossing-point at $19.6~K$ and a high field crossing 
point at $22.7~K$  Presenting these data on the $T_{cr}~vs.~\mu _oH$ plot 
of Fig. 4 shows that   
the $x=0.081$ sample (open squares) has a 
plateau from about $0.3$ to $1.0~T$, and it has a second plateau from about $3$ to $7~T$ 
with a gradual transition from $1-3~T$.  Here, data on the low field plateau fit best to $2D$ scaling, and data on the high field 
plateau fit best to $3D$ scaling as shown in Fig. 5a and Fig. 5b.  In a similar manner, data for the $x=0.070$ sample (solid circles in Fig. 4) shows both a low field plateau at $14.2~K$ and a high field plateau at $15.2~K$.  
Here again, the low field plateau data fit $2D$ scaling and the high field plateau data 
fit $3D$ scaling as shown by Fig. 5c and Fig. 5d.  Both these samples show that with an increasing magnetic field, the 
sample undergoes a cross-over from $2D$ scaling to $3D$ scaling as $H_{c2}$ is approached.

To be a bit more quantitative about the $2D$ to $3D$ cross-over, we plot the Rosenstein et al. parameters\cite {6}vs. magnetic field.  Within this model, if

$$\xi _c/d ~\ll~1/2[\sqrt {(b+t-1)^2+4bt}+(b+t-1)]=f^{2D},~\eqno (1)$$

then one expects $2D$ scaling.  Here, $\xi _c$ is the c-axis coherence distance, $d$ is the copper oxide plane spacing, $b=H/H_{c2}(0)$, and $t=T/T_{co}$.  Similarly, within this model, if

$$\xi _c/d~\gg~b+t-1+[2(b+t-1)^3]/27~=f^{3D},~\eqno (2)$$

then one expects $3D$ scaling.  It should be pointed out that\cite {6} 

$$b+t-1~=~[T-T_c(H)]/T_c.$$

so this variable measures the reduced temperature interval from the $H_{c2}$ line.

Both $f^{2D}$ and $f^{3D}$ are plotted vs. $\mu _oH$ for each crossing-point, $T_{cr}$ 
in the insets of Fig. 4 for the $x=0.081$ and $0.070$ samples. From the plot of $f^{2D}$ 
(open triangles in both insets), it is clear that the region of $3D$ scaling  
begins at a magnetic field where $f^{2D}$ 
becomes as large as 1.0.  The $f^{3D}$ functions are also plotted for completeness.   
These magnetization data do not provide a measure of $\xi _c$, but it is of interest to note 
that $3D$ scaling begins at a magnetic field where the Rosenstein et al. $f^{2D}$ 
function\cite {6}reaches $1.0$ for both $x=0.081$ and $x=0.070$ samples.

\section{Conclusions}

The $La_{2-x}Sr_{x}CuO_{4+\delta }$ high temperature superconductor system provides 
a rich arena to study vortex fluctuations.   Near optimum doping the $x=0.143$ and 
$x=0.156$ samples show $3D$ fluctuations over the entire magnetic field range studied.  The 
c-axis coupling is strong enough in the optimally doped samples 
to make the fluctations $3D$ at all the measured fields.  Stated another way, $\xi _c/d$ is greater than one for these samples.    
As discussed previously\cite {8} a reduction of carrier concentration tends to make the material 
more $2D$ at magnetic fields substantially below $H_{c2}(0)$, presumably because the 
relatively weak c-axis coupling  in these cuprates is made even weaker with depleted carrier 
concentration.
Reduction of the doping to $x=0.117$ and $x=0.10$ gave samples that showed $2D$ 
fluctuations for all magnetic fields up to $7~T$.  Presumably the $f^{2D}$ did not get large enough for fields up to $7~T$ in these two samples to cause a $2D$ to $3D$ transition.  For all four of these samples, $H_{c2}(0)$ 
was over $30~T$ so the $b=H/H_{c2}(0)$ values are small compared to the reduced temperature, 
$t=T/T_c$  terms and thus the $b$ term in Eq. (1) does not make a large enough contribution 
to cause the cross-over.
With further reduction in doping to $x=0.081$ and $x=0.070$, the data probes fields comparable 
to $H_{c2}(0)$.  Then the $b=H/H_{c2}(0)$ term in Eq. (1) is important,  as shown in the inset of Fig. 4, and a clear cross-over from $2D$ to $3D$ behavior was seen.  

The YBCO and $La_{2-x}Sr_xCuO_{4+\delta }$ data differ in one important respect.  
The YBCO data\cite {5,6}show $3D$ behavior at lower fields and $2D$ behavior at higher fields, whereas these  $La_{2-x}Sr_xCuO_{4+\delta }$ data show $2D$ behavior at low field and $3D$ 
behavior at high field.  The authors do not understand this difference, but in light of Eq. (1) and 
 the data in 
the inset in Fig. 4, the $2D$ to $3D$ cross-over with increasing field  seems to be a reasonable 
result.

\section{Acknowledgments}
Comments from V. G. Kogan and J. R. Clem were very helpful.  
Ames Laboratory is operated for the
U. S. Department of Energy by Iowa State University under contract No.
W-7405-ENG-82 and
supported by the DOE, the Office of Basic Energy Sciences.

\begin{figure}
\caption{Magnetization vs temperature plot for an $x=0.143$ sample near optimum doping.  
The crossing point changes as the field increases from $0.95~T$ to $2.0~T$.}
\end{figure}

\begin{figure}
\caption{Magentization crossing points for two successive fields, $T_{cr}$, as a 
function of field.} 
\end{figure}

\begin{figure}
\caption{$4\pi M~vs.~T$ data for the $x=0.081$ sample showing two distinct crossing-points 
for the low field and high field regions. }
\end{figure}

\begin{figure}
\caption{Comparison of the plateau regions for $T_{cr}$ with the values of the 
theoretical $f^{2D}$ and $f^{3D}$ parameters for the (a) $x=0.081$ and (b) $x=0.070$ 
samples.}
\end{figure}

\begin{figure} 
\caption{$2D$ and $3D$ scaling plots for the two plateaus of the $x=0.081$ (a, b) and 
the $x=0.070$ samples (c ,d).}
\end{figure}

\vfil\eject

\end{document}